\documentclass{KapProc} % Computer Modern font calls
\usepackage{graphicx}
\newcommand{\beq}{\begin{equation}}
\newcommand{\eeq}{\end{equation}}
\newcommand{\ben}{\begin{eqnarray}}
\newcommand{\een}{\end{eqnarray}}
\newcommand{\bea}{\begin{array}}
\newcommand{\eea}{\end{array}}

\newcommand{\bef}{\begin{figure}}
\newcommand{\eef}{\end{figure}}

\normallatexbib

\begin{document}
\articletitle{ Bunching and antibunching from single NV color
centers in diamond}

\author{A. Beveratos, R. Brouri, J.-P. Poizat, P. Grangier}
\affil{Laboratoire Charles Fabry de l'Institut  d'Optique, UMR 8501 du CNRS, \\
B.P. 147, F91403 Orsay Cedex - France}
\email{jean-philippe.poizat@iota.u-psud.fr}

%\preprint{Version IOTA, \today}
%\draft

\begin{abstract}

We investigate correlations between fluorescence photons emitted
by single N-V centers in diamond with respect to the optical
excitation power. The autocorrelation function shows clear photon
antibunching at short times, proving the uniqueness of the
emitting center. We also report on a photon bunching effect, which
involves a trapping level. An analysis  using rate-equations for
the populations of the N-V center levels  shows the intensity
dependence of the rate equation coefficients.

\vspace{0.5cm}

\vspace{0.5cm}

\end{abstract}

\section*{Introduction}

Quantum cryptography relies on the fact that single quantum states
can not be cloned. In this way coding information on single
photons would ensure a secure transmission of encryption keys.
First attempts for quantum cryptography systems \cite{MBG,TRT}
were based on attenuated laser pulses. Such sources can actually
produce isolated single photons, but the poissonian distribution
of the photon number does not  guarantee both the uniqueness of
the emitted photon and a high bit-rate. Therefore a key milestone
for efficient and secure quantum cryptography systems is the
development of single photon sources.

Several pioneering experiments have already been realized in order
to obtain single photon sources. Among these attempts one can cite
twin-photons experiments \cite{GRA,HM}, coulomb blockade of
electrons in quantum confined heterojunctions \cite{KBKY}, or
fluorescence emission from single molecules \cite{MGM,BLTO}. Up to
now interesting results were obtained, but considerable  work is
still needed to design a reliable system, working at room
temperature with a good stability and well-controlled emission
properties. More recently several systems have been proven to be
possible candidates for single photon sources. Antibunching was
indeed observed in CdSe nanospheres \cite{MIMCSB}, thereby proving
the purely quantum nature of the light emitted by these sources.
In the same way photon antibunching in Nitrogen-Vacancy (N-V)
colored centers in diamond was reported by our group \cite{BBPG}
and by Kurtsiefer et al \cite{KMZHW}. A remarkable property of
these centers is that they do not photobleach at room temperature:
the fluorescence level remains unchanged after several hours of
continuous laser irradiation of a single center in the saturation
regime. These centers are therefore very promising candidates for
single photon sources.

In this paper we present further investigation of the fluorescence
light emitted by NV colored centers. We analyze the system dynamic
through the measurement of the autocorrelation function, showing
the existence of a shelving effect which reduces the counting rate
of the fluorescence emission, and gives rise to photon bunching
\cite{BFTO}. We study the dependence of this behavior on the
pumping power, and we compare the experimental results with the
predictions of a three-level model using rate equations.

\section{Experimental setup}

Our experimental set-up is based on a home-made scanning confocal
microscope.

\begin{figure}[!ht]
\center
\includegraphics[scale=0.37]{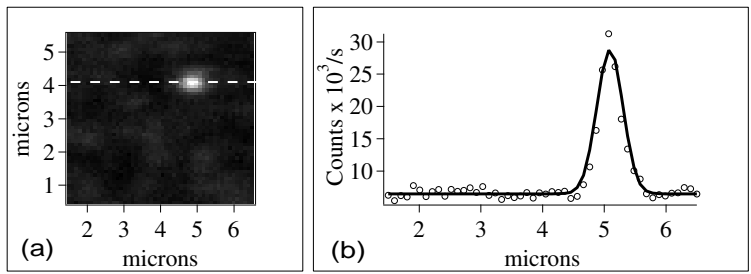}
\caption{ (a) Confocal microscopy raster scan ($ 5 \; \times \;  5
\; \mu$m$^2$) of the sample performed about $10  \; \mu m$ below
the diamond surface. The size of a pixel is $100$ nm. The
integration time per pixel is $32$ ms. The laser intensity
impinging on the sample is $15$ mW . (b) Line scan along the
dotted line. The data is shown together with a gaussian fit, which
is used to evaluate the signal and background levels (right). Here
we obtain $\rho= S/(S+B) = 0.81$} \label{Scans}
\end{figure}
\pagebreak

A CW frequency doubled Nd:YAG laser ($\lambda = 532 nm$) is
focused on the sample by a high numerical aperture ($1.3$)
immersion objective. A PZT-mounted mirror allows a x-y scan of the
sample, and a fine z-scan is obtained using another PZT.

The sample is a $0.1 \times 1.5 \times 1.5 mm^3 [110]$ crystal of
synthetic Ib diamond from Drukker International. The
Nitrogen-Vacancy centers consist in a substitutional nitrogen with
an adjacent vacancy, and are found with a density of about 1
$\mu$m$^{-3}$ in Ib diamond. The centers can be seen with a signal
to background ratio of about $5$ by scanning the sample as shown
in figure \ref{Scans}, and a computer-controlled servo-loop allows
to focus on one center for hours.

The fluorescence is collected by the same objective, and separated
from excitation light by a dichroic mirror. High rejection
($10^{10}$) high-pass filters remove any leftover pump light.
Spatial filtering is achieved by focusing on a $50 \mu m$ pinhole.
The fluorescence is then analyzed by an ordinary Hanbury-Brown and
Twiss set-up. The time delay between the two photons is converted
by a time-to-amplitude converter (TAC) into a voltage which is
digitalized by a computer board.
%(National Instruments PCI-1200).
\begin{figure}[!ht]
\center
\includegraphics[scale=0.4]{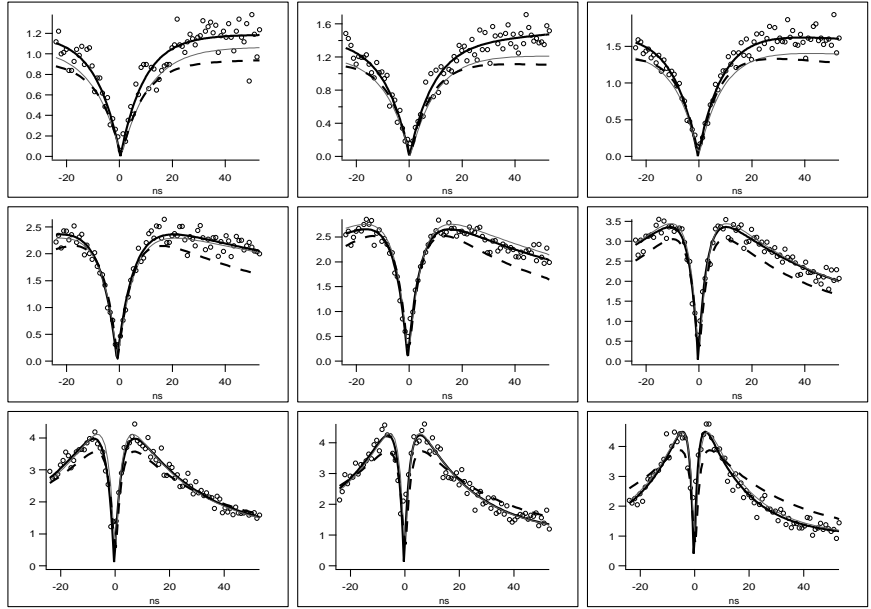}
\caption{Experimental data of $g^{(2)}(\tau)$ for different values
from 0.3 to 31 mW of pumping power. The solid line represents the
fit. The dashed (resp. thin) lines are calculations of
$g^{(2)}(\tau)$ for $k_{23}$ and $k_{32}$ (see figure 3 for
definition) not depending (resp. depending) on pump
power. Details are given in the text of section 2.} \label
{Experimental results}
\end{figure}

\pagebreak In a low counting regime, this system allows us to
record the autocorrelation function:
$g^{(2)}(\tau)=\frac{<I(t)I(t+\tau )>}{<I(t)>^2}$.

Experimental data were corrected from background noise, and
normalized to the coincidence number corresponding to a poissonian
source of equivalent power \cite{BBPG}. Different correlation
functions obtained for several pumping power are shown in figure
\ref{Experimental results}. These functions are centered around
$\tau=0$, but one can notice slight drifts attributed to thermal
drift of our acquisition electronics. For $\tau=0$,
$g^{(2)}(\tau)$ goes clearly down to zero, which is the signature
of a single emitting dipole. On the other hand, $g^{(2)}(\tau)$
goes beyond 1 at longer times, and then decays to 1. This behavior
is an evidence of the presence of a trapping level. To correctly
describe the system we therefore considered a 3-level system.

\section{Theoretical background and discussion}

\begin{figure}[!ht]
\center
\includegraphics[scale=0.4]{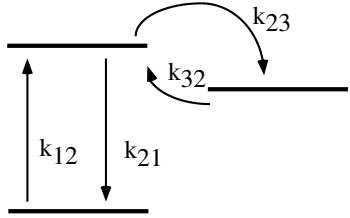}
\caption{Three level system as used for modelisation}
\label{System3n}
\end{figure}

Owing to the fast damping of coherences, we use rate equations.
Let us consider the 3-level scheme described in figure \ref
{System3n}.
The evolution of the populations are therefore given by: \beq
\label{populations} \frac{d}{dt} \left(
\begin{array}{c}
                                        \sigma_1 \\
                                        \sigma_2 \\
                                        \sigma_3
    \end{array} \right) = \left( \begin{array}{ccc}
                                    k_{12} & k_{21} & 0 \\
                                    k_{12} & -k_{21}-k_{23} & k_{32} \\
                                    0 & k_{23} & -k_{32}
    \end{array} \right)
    \left( \begin{array}{c}
                                        \sigma_1 \\
                                        \sigma_2 \\
                                        \sigma_3
    \end{array} \right)
\eeq
 with the initial condition $\sigma_1 =1$, $\sigma_2 =\sigma_3
=0$ at $t=0$, which means that a photon has just been emitted and
the system is therefore prepared in its ground state. The decay
rate from level 3 to 1 is neglected \cite{DFTJKNW}.

\begin{figure}[!ht]
\center
\includegraphics[scale=0.31]{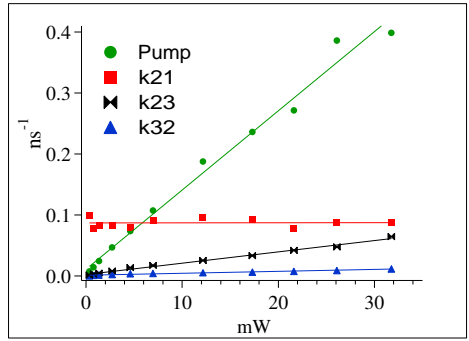}
\caption {Evolution of $k_{ij}$ with respect to the pump power}
\label{kij}
\end{figure}

By analytically solving equations \ref{populations}, one can
derive \beq \label{fitg2}
 g^{(2)}(\tau) = \frac{\sigma_{2} (\tau )}{\sigma_{2} (\infty)} =
  1- \frac{1+g_e}{2} \exp^{ \frac{k_{tm}+k_{1m}}{2}
\tau} - \frac{1-g_e}{2} \exp^{- \frac{k_{tm}-k_{1m}}{2} \tau} \eeq
with the stationary population:
 \beq \label{saturation}
 \sigma_2 (\infty) = {\sigma_{2}}_{\infty} = \frac{k_{12}k_{32}} {k_{12}
 k_{23}+ k_{12}k_{32}+k_{21}k_{32}} \eeq
and with
 \beq k_{tm} = k_{12}+ k_{21}+ k_{23}+ k_{32} \eeq \beq
k_{1m}^{2}=\left( \left( k_{12}+ k_{21}- k_{23}- k_{32} \right) ^{2} \\
                                             + 4k_{21}k_{23} \right)
\eeq \beq g_{e}=\frac {2k_{12}k_{23}+k_{32} \left( k_{12}+
k_{21}- k_{23}- k_{32} \right)}
                        {k_{1m}k_{32}}
\eeq We thus have four equations which enables us to express the
four rates $k_{ij}$ with respect to the experimental variables
$g_{e} , k_{tm} , k_{1m} , {{\sigma_{2}}_{\infty}}$. By fitting
the experimental values of $g^{(2)}$ with expression \ref{fitg2}
we obtain the values of $g_{e}$, $k_{tm}$ and $k_{1m}$ for each
value of the intensity. The last value needed in order to solve
the system concerns the parameter ${\sigma_{2}}_{\infty}$ which is
directly linked to the count rate \beq \label{equN} N=\eta \times
k_{21} \times \sigma_{2 _{\infty}} \eeq We assume that $k_{21}$
doesn't depend on the pump power, and we set $\eta$ to the value
for which this condition is satisfied. We found $\eta = 3 \times
10^{-3}$ which is in perfect agreement with our estimated value
\cite{BBPG}. The parameters $k_{ij}$  are plotted as functions of
pumping power in figure \ref{kij}.

Note that $k_{12}$ is, as expected, a linear function of the
pumping power, and that $k_{21}^{-1}$ has a constant value of
11.6ns, which corresponds to results reported in literature
\cite{CTJ}.

What is noteworthy here is that $k_{23}$ and $k_{32}$ linearly
depend on pump power, with $k_{23}$ greater than $k_{32}$, which
means that at high pumping power the system tends to be shelved
into the third level. This intensity-dependent effect has not been
reported yet. In Figure \ref {Experimental results}, three plots
were superimposed to the experimental results. The solid line
represents the best fit. The dashed line represents the result of
equation \ref{fitg2} with values of $k_{23}$ and $k_{32}$
independent on intensity taken from reference \cite{KMZHW}. One
can see in this last case that the agreement between calculated
and experimental value is not fully satisfactory. Finally, the
thin line is
 obtained by using the values of
$k_{23}$ and $k_{32}$ given by the linear fit of  figure
\ref{kij}. Agreement  with the experimental results is good.

% Assuming a linear dependence of $k_{23}$
%and $k_{32}$ on intensity leads however to the results presented
%with a thin line in figure \ref {Experimental results}, which
%correctly account for experimental results.

This intensity dependence of $k_{23}$ and $k_{32}$ can also be
observed on the total photon counts. In fact the number of photons
emitted per second should decrease as the trapping in the
metastable state increases.
\begin{figure}[!ht]
\center
\includegraphics[scale=0.31]{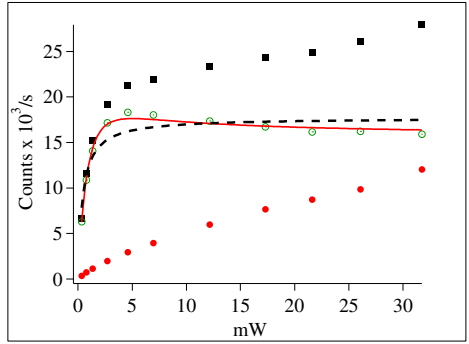}
\caption{Saturation behaviour of the N-V center. The solid line
gives the fit by using our model, and the dashed line the results
for normal saturation behaviour. Values are plotted in respect to
pumping power. The intensity may be deduced from the focused beam
size, which in our case is not accurately known due to spherical
aberrations.} \label{graph_saturation}
\end{figure}
Figure \ref{graph_saturation} shows the count rate of the single
N-V center as a function of the pump power. We clearly see a
decrease in the photon counts. We can fit with a good agreement
our experimental data using equations \ref{equN} and
\ref{saturation}.

We have repeated this experiment with the $532 nm$ Nd:YAG line and
the $514 nm$ Argon line and found similar linear dependency of the
$k_{23}$ and $k_{32}$ rates with the pump power.

\section*{Conclusion}

We have measured the autocorrelation function of the fluorescence
light emitted from a single NV colored center in diamond using a
confocal microscope. Our results are in very good agreement with
what can be expected from rate equations in a 3-level scheme,
provided that a linear dependence of some rate coefficients with
pumping power is assumed. Such an effect has not been reported
yet, and must be taken into account when designing a pulsed
single-photon source.

\pagebreak

%%%%%%%%%%%%%%%%%%%%%%%%%%%%%%%%%%%%%%%%%%%%%%%%%%%%%%%%%%%%%%%%
\end{document}